\documentclass[twocolumn,showpacs,preprintnumbers,amsmath,amssymb,superscriptaddress]{revtex4-1}
\usepackage{graphicx}
\usepackage{dcolumn}
\usepackage{bm}
\usepackage{CJK, CJKnumb}
\usepackage{color}
\begin{document}
\begin{CJK*}{GBK}{song}

\title{Superconducting lens and Josephson effect in AA-stacked bilayer graphene}
\author{Wei-Tao Lu} \email{physlu@163.com}
\affiliation{School of Physical Science and Technology, Nantong University, Nantong 226019, China}
\author{Tie-Feng Fang}
\affiliation{School of Physical Science and Technology, Nantong University, Nantong 226019, China}
\author{Qing-Feng Sun} \email{sunqf@pku.edu.cn}
\affiliation{International Center for Quantum Materials, School of Physics, Peking University, Beijing 100871, China}
\affiliation{Hefei National Laboratory, Hefei 230088, China}

\begin{abstract}
We study the superconducting transport phenomena, involving lensing effect and supercurrent in AA-stacked bilayer graphene, which is characterized by a linear gapless band with two shifted Dirac cones.
Our findings indicate that cross Andreev reflection and Josephson current occur exclusively within the intracone process, while intercone scatterings are strictly prohibited.
The normal/superconductor/normal junction can act as a superconducting lens for the upper and lower cones.
Depending on cone index, the transmitted electrons and holes can be focused, collimated or diverged by adjusting the gate voltages.
In superconductor/normal/superconductor junction, due to interlayer coupling, the critical currents of the two cones exhibit distinct oscillation periods with junction width, leading to an irregular oscillation of the total critical current.
Furthermore, the oscillations of critical currents with exchange field maintain a stable phase difference of one quarter period between the two cones.
Consequently, a cone-dependent $0-\pi$ transition is achieved in this Josephson junction.
\end{abstract}
\maketitle

\section{Introduction}

Graphene-superconductor interfaces exhibit a variety of intriguing Andreev processes \cite{Beenakker}, including crossed Andreev reflection (CAR) \cite{Cayssol, Sun, addr1, Linder, Pandey, Sun2}, 
superconducting lens \cite{Gomez, Cheraghchi, addr2}, 
and Josephson effect \cite{Titov, Sun3, Linder2, Sun4, Wakamura, Pellegrino, Wang, addr3}, 
which have been predicted to occur under different conditions and hold applications in superconducting electronic devices.
Due to the linear energy dispersion of monolayer graphene (MLG), an electron beam through the $npn$ junction experiences negative refraction which may act as a Veselago lens for electrons \cite{Cheianov, Moghaddam}, and experimental evidence supporting this phenomenon has been reported \cite{GHLee}.
This observation stimulates further study on the superconducting hybrid systems \cite{Gomez, Cheraghchi, Wambaugh}.
In the MLG-based normal/superconductor/normal (N/S/N) junction, both propagating electron and hole waves could be focused into distinct points by virtue of CAR \cite{Gomez}.
On the other hand, the anomalous scaling behavior regarding supercurrent is demonstrated in superconductor/normal/superconductor (S/N/S) MLG setup \cite{Titov}.
As the proximity-induced exchange splitting appears, the critical current at the $0-\pi$ transition remains finite and significantly exceeds that observed in metallic systems \cite{Linder2}.
Spin-orbit interactions may enhance the robustness of supercurrent through MLG Josephson junctions \cite{Wakamura}, while dilute impurities could diminish the supercurrent \cite{Pellegrino}.

Recently, there has been growing attention on the superconducting transport properties in bilayer graphene (BLG), such as Bernal BLG \cite{Ludwig, Efetov, Soori, Xie, PRam, Park, Rout, addr4} 
and twisted BLG \cite{addr3, Xie2, Alvarado, Diez, Sainz, Hu}.
BLG is regarded as a more suitable system compared to MLG for observing Andreev reflection (AR) \cite{Ludwig, Efetov, Soori}.
The mode number of Andreev bound states (ABS) in Bernal BLG Josephson junctions can be modulated by the superconducting coherence length, allowing one to estimate the supercurrent quantitatively \cite{Park}.
In twisted BLG, the valley-polarized state \cite{Xie2} and chiral pairing \cite{Alvarado} would give rise to $\phi_0$-Josephson junctions.
Interestingly, the superconducting diode effect in twisted BLG is realized \cite{Diez, Hu}, where the critical current magnitudes differ when the currents flow in opposite directions.

The electronic properties of BLG depend sensitively on the stacking sequence.
Unlike Bernal BLG and twisted BLG, the AA-stacked BLG (AA-BLG) has a linear dispersion, but the two resultant cones are split in energy, owing to the interlayer coupling \cite{Rozhkov}.
Recent advancements in growing stable AA-BLG \cite{JKLee, ZLiu} open the study on the properties of this class of graphene.
Numerous studies have reported that AA-BLG shows many intriguing properties, including electronic transport \cite{Tabert, Sanderson, Abdullah, Yang}, antiferromagnetism \cite{Sboychakov, Apinyan}, phase transition \cite{Sboychakov2, Apinyan2, YLi, Georgoulea}, photogalvanic effects \cite{Zheng}, and quantum dots \cite{Qasem, Rakhmanov}.
A salient feature of AA-BLG is that the two Dirac cones coincide at Fermi surfaces [see Fig. 1(b)], instead of Fermi points in MLG.
Especially, it is demonstrated that the domain wall between MLG and AA-BLG can generate two distinct types of collimated beams corresponding to the lower and upper cones \cite{Abdullah}.
When AA-BLG is twisted relative to an MLG at certain angles, the two Dirac cones will become anisotropic and multiple \cite{YLi}.
Up to now, however, few works have paid attention to the AA-BLG-based superconductor junctions \cite{Alidoust, WTLu}.
Preliminary findings suggest that by modulating the gate voltages, the local AR can manifest as either specular or retroreflective AR according to the cone index \cite{WTLu}.
Nonetheless, phenomena such as superconducting lens and Josephson effect in AA-BLG warrant further exploration, and the effect of cone degree of freedom remains unresolved.

This paper is dedicated to the theoretical study of the Dirac cone-related superconducting lens and Josephson current in AA-BLG.
Because of the two Dirac cones in its energy band, AA-BLG offers a unique opportunity for the manifestation of superconducting lens and Josephson effect.
It is found that by adjusting the value of chemical potential relative to the interlayer coupling, the N/S/N junction shows various lens effects depending on the cone index, which can be detected via conductance at the focal spots.
In S/N/S junction, the critical currents of the two cones exhibit different oscillation behaviors as functions of the junction width and the exchange field.
Notably, the $0-\pi$ transition and switch effect for a certain cone are achievable.

The organization of this paper is as follows.
In Sec. II, we give the Hamiltonian of the proposed junctions, and derive the Andreev conductance and ABS for the two cones.
Section III is devoted to numerical results on superconducting lens and Josephson current.
Finally, a brief summary is presented in Sec. IV.

\section{Theoretical Formulation}

AA-BLG consists of two MLG layers and every carbon atom $A$ ($B$) of the top layer $1$ is stacked directly above the corresponding atom $A$ ($B$) of the bottom layer $2$ with a direct interlayer coupling $\gamma$ \cite{Lobato}.
In the basis $\Psi=(\psi_{A1},\psi_{B1},\psi_{A2},\psi_{B2})^T$ with elements referring to the sublattices in each layer, the matrix form of AA-BLG's Hamiltonian within the continuum approximation in the low energy regime can be written as \cite{Tabert, Sanderson, Abdullah}
\begin{align}
H_{\eta s} = \left(\begin{array}{cc} H_0 - \mu - s h  &  \gamma \sigma_0  \\  \gamma \sigma_0 & H_0 - \mu - s h \end{array}\right), \label{eq1}
\end{align}
and $H_0=\hbar v_F (k_x \sigma_x + \eta k_y \sigma_y)$ is the Hamiltonian of MLG.
$\sigma=(\sigma_x, \sigma_y)$ are the Pauli matrices in the sublattice spaces and $\sigma_0$ is $2\times 2$ unit matrix.
$(k_x, k_y)$ denote the wave vectors. $v_F$ is the Fermi velocity.
The indexes $\eta = \pm 1$ and $s=\pm 1$ are for the valleys $K/K'$ and spin up/down, respectively.
$\mu$ is chemical potential controlled by the gate voltages.
The exchange field $h$ can be induced via the proximity effect by depositing the magnetic insulator, such as EuS, on AA-BLG samples \cite{PWei}.

We consider the AA-BLG-based N/S/N and S/N/S models in the $x-y$ plane, which vary along the $x$ direction, as shown in Fig. 1(a).
The superconductivity in AA-BLG can be generated by the superconducting proximity effect \cite{Heersche}.
The electron and hole excitations are described by the Bogoliubov-de Gennes (BdG) equation
\begin{align}
H_{BdG}\left(\begin{array}{cc} u \\ \nu \end{array}\right)=\left(\begin{array}{cc} H_{\eta s}  &  \Delta(T) e^{i \phi}  \\  \Delta(T) e^{i \phi} & -H_{\eta \bar{s}} \end{array}\right) \left(\begin{array}{cc} u \\ \nu \end{array}\right) = \epsilon \left(\begin{array}{cc} u \\ \nu \end{array}\right) \label{eq2}
\end{align}
where $\bar{s}=-s$, $\phi$ is the superconducting phase, and $u$ (or $\nu$) is the electron (or hole) wave function.
The pair potential is $\Delta(T)=\Delta_0 \tanh(1.74 \sqrt { \frac{T_c}{T} -1} )$ with the zero-temperature superconducting gap $\Delta_0$ and the transition temperature $T_c = 0.57 \Delta_0 / k_B$.
The dispersion relations can be obtained by the eigenvalues of BdG Hamiltonian, which are $\epsilon_e=-\mu_N - s h \pm \sqrt{k_x^2+k_y^2} + \tau \gamma$ and $\epsilon_h=\mu_N - s h \pm \sqrt{k_x^2+k_y^2} - \tau \gamma$ for electron and hole in the N region ($\Delta_0=0$), respectively. 
Here, $\tau=\pm$ are for the upper and lower cones.
The energy bands for AA-BLG in the normal phase and superconducting phase are plotted in Fig. 1(b).
We can see that the linear band consists of two Dirac cones, namely, lower cone $\epsilon_-$ and upper cone $\epsilon_+$, shifted by $2\gamma$.
This provides physical mechanism for specific signatures of Andreev scattering in AA-BLG.
Energy band in the S region has the form $\epsilon_S=\pm \sqrt{(\mu_S + \tau \gamma \pm \hbar \sqrt{k_x^2+k_y^2})^2 + \Delta^2}$.
The potential $\mu$ is labeled as $\mu_N$ (or $\mu_{L, R}$) and $\mu_S$ in the N and S regions, respectively.
The exchange field $h$ is only applied in the N region of S/N/S junction.
The eigenstates of BdG Hamiltonian are given in Appendix A.
Note that the effect of interface $\delta$ barrier is not considered due to the linear dispersion relation and Klein tunneling of graphene \cite{Beenakker}.
Owing to the valley degeneracy in the proposed systems, herein we only discuss the transport of one valley.

Considering an N/S/N junction with the width of the center S region being $d$ and the SN interfaces at $x=0$, $d$, for an incident electron from Dirac cone $\tau$, the wave functions in the left N, center S, and right N regions are
\begin{align}
\psi_L=&\phi_{\tau e L}^+ + r_e^\tau \phi_{\tau e L}^- + r_e^{\bar{\tau}} \phi_{\bar{\tau} e L}^- + r_h^\tau \phi_{\tau h L}^- + r_h^{\bar{\tau}} \phi_{\bar{\tau} h L}^-, \\
\psi_S=&a_1 \psi_{+S1}^+ + a_2 \psi_{+S1}^- + a_3 \psi_{-S1}^+ + a_4 \psi_{-S1}^-  +  b_1 \psi_{+S2}^+    \nonumber   \\
&+ b_2 \psi_{+S2}^- + b_3 \psi_{-S2}^+ + b_4 \psi_{-S2}^-,  \\
\psi_R=&t_e^\tau \phi_{\tau e R}^+ + t_e^{\bar{\tau}} \phi_{\bar{\tau} e R}^+ + t_h^\tau \phi_{\tau h R}^+ + t_h^{\bar{\tau}} \phi_{\bar{\tau} h R}^+,
\end{align}
with $\bar{\tau}=-\tau$. The states $\phi_{\tau e L/R}^\pm$ and $\phi_{\tau h L/R}^\pm$ are, respectively, the eigenstates for electron and hole in the left/right N regions (see Appendix A for detail).
The coefficients $r_e^\tau$, $r_h^\tau$, $t_e^\tau$, and $t_h^\tau$ correspond to normal reflection, local AR, elastic tunneling (ET), and CAR, respectively.
CAR is a nonlocal AR that has applications in generating entangled states and designing Cooper pair splitters \cite{Beenakker, Pandey}.
The coefficients $r_e^{\tau,\bar{\tau}}$, $r_h^{\tau,\bar{\tau}}$, $t_e^{\tau,\bar{\tau}}$, $t_h^{\tau,\bar{\tau}}$, and the constants $a_{1,2,3,4}$ and $b_{1,2,3,4}$ can be obtained from the continuity of wave functions at the SN interfaces \cite{addr5}.
Then the corresponding reflection and transmission probabilities can be derived from the probability current.
For the considered system, the probability current density operator is defined as 
\begin{align}
\mathcal{J}=\frac{-i}{\hbar} [r, H_{BdG}]=v_F \tau_z \otimes \sigma_0 \otimes (\sigma_x \mathbf{e}_x + \eta \sigma_y \mathbf{e}_y), 
\end{align}
where $\tau_z$ is Pauli matrix denoting electron-hole index and $r=x \mathbf{e}_x + y \mathbf{e}_y$ with the unit vectors $\mathbf{e}_{x/y}$ in $x/y$ directions. 
Only the $x$ component of the current density operator $\mathcal{J}_x=v_F \tau_z \otimes \sigma_0 \otimes \sigma_x$ decides the reflection and transmission probabilities at the SN interface, which can be obtained straightforwardly: 
\begin{align}
&R_e^\tau= \Big| \frac{\langle \phi_{\tau e L}^- | \mathcal{J}_x | \phi_{\tau e L}^- \rangle }{\langle \phi_{\tau e L}^+ | \mathcal{J}_x | \phi_{\tau e L}^+ \rangle }\Big| |r_e^\tau|^2,  \label{eqS1}
\end{align}
\begin{align}
&R_h^\tau= \Big| \frac{\langle \phi_{\tau h L}^- | \mathcal{J}_x | \phi_{\tau h L}^- \rangle }{\langle \phi_{\tau e L}^+ | \mathcal{J}_x | \phi_{\tau e L}^+ \rangle }\Big| |r_h^\tau|^2,  \label{eqS2}  \\
&T_e^\tau=\Big| \frac{\langle \phi_{\tau e R}^+ | \mathcal{J}_x | \phi_{\tau e R}^+ \rangle }{\langle \phi_{\tau e L}^+ | \mathcal{J}_x | \phi_{\tau e L}^+ \rangle }\Big| |t_e^\tau|^2,  \label{eqS3}  \\
&T_h^\tau=\Big| \frac{\langle \phi_{\tau h R}^+ | \mathcal{J}_x | \phi_{\tau h R}^+ \rangle }{\langle \phi_{\tau e L}^+ | \mathcal{J}_x | \phi_{\tau e L}^+ \rangle }\Big| |t_h^\tau|^2.   \label{eqS4}
\end{align}
Note that owing to the orthogonality of the eigenstates between the two cones \cite{Sanderson}, the intercone AR scattering is forbidden and only intracone process is allowed.
For an incident electron from cone $\tau$, the Andreev scatterings satisfy $R_e^{\bar{\tau}} = R_h^{\bar{\tau}} = T_e^{\bar{\tau}} = T_h^{\bar{\tau}} =0$ and the conservation condition
\begin{align}
R_e^\tau + R_h^\tau + T_e^\tau + T_h^\tau =1.
\end{align}

Then, the conductances $G_{CAR}^\tau$ and $G_{ET}^\tau$ carried by CAR and ET can be evaluated using the Blonder-Tinkham-Klapwijk formula \cite{Blonder}
\begin{align}
G_{CAR, ET}^\tau = G_0 \int_{-\pi/2}^{\pi/2} T_{h, e}^\tau \cos \alpha_i d \alpha_i,
\end{align}
where $\alpha_i$ is the incident angle with respect to the $x$ direction, $G_0=e^2 W |k_e^\tau| / \pi h$ characterizes the ballistic conductance of the N/S/N junction, and $W$ labels the width of the junction in $y$ direction.
When both S region and left N lead are grounded, while a voltage $V_R$ is applied to the right N lead, the current from the left N lead flowing into the superconductor can be written as $I_L^\tau=(G_{CAR}^\tau-G_{ET}^\tau) V_R$ at the low temperature \cite{Sun5}.
Then, the cone-dependent nonlocal differential conductances is obtained, $G^\tau = d I_L^\tau / d V_R = G_{CAR}^\tau- G_{ET}^\tau$.
The differential conductance $G^\tau$ is positive when $G_{CAR}^\tau$ dominates and negative when $G_{ET}^\tau$ dominates, meaning that the current $I_L^\tau$ is increased or decreased as the voltage $V_R$ increases.
Note that the subscripts $e$ and $h$ in Eqs. (\ref{eqS1})-(\ref{eqS4}) are for the electron and hole components in the Nambu space.
The energy bands of the electron and the hole are symmetric with respect to Fermi energy $E_F$, and their contribution to the current is the same \cite{Gomez}.
In the subsequent discussion, we take an electron as the incident quasiparticle, and the scenarios arising from an incident hole can be analyzed analogously.

For the S/N/S Josephson junction, the supercurrent across the junction between two superconductors is mainly contributed by the ABS.
We set $\phi_{L/R}=\mp \phi/2$ in the left/right S regions and the SN interfaces are at $x=\pm w/2$ with $w$ the width of the center N region.
So $\phi=\phi_R - \phi_L$ is the phase difference.
The continuity of wave functions relates states at the left and right interfaces by $u(w/2)=M_e u(-w/2)$ and $\nu (w/2)=M_h \nu (-w/2)$.
Wave functions for electron and hole satisfy $\nu (-w/2)=M_S u(-w/2)$ and $\nu (w/2)=M_S^{-1} u (w/2)$ at the two interfaces.
Here, $M_e$, $M_h$, and $M_S$ are the complicated $4 \times 4$ matrices (not shown here) calculated from Eqs. (\ref{eq22})-(\ref{eq25}).
As a consequence, we yield the relation $u (-w/2) = M_e^{-1} M_S M_h M_S u (-w/2)$.
The existence of ABS requires
\begin{align}
Det (I - M_e^{-1} M_S M_h M_S) = 0, \label{eq10}
\end{align}
based on which, one can get the product formula
\begin{align}
\prod_\tau \frac{\cos \phi - A_\tau \cos 2 \beta - B_\tau \sin 2 \beta - C_\tau}{\cos \alpha_e^\tau \cos \alpha_h^\tau}=0. \label{eq11}
\end{align}
The parameters $\beta$, $\alpha_e^\tau$, and $\alpha_h^\tau$ are given in Appendix A.
The formulations of $A_\tau$, $B_\tau$, and $C_\tau$ can be found in Appendix B.
Equation (\ref{eq11}) 
suggests that the energy levels of ABS between the two cones are irrelevant, and so the supercurrent just occurs in the intracone.
Consequently, Eq. (\ref{eq11}) for both cones can be rewritten as
\begin{align}
\cos \phi = A_\tau \cos 2 \beta + B_\tau \sin 2 \beta + C_\tau. \label{eq12}
\end{align}
For a heavy doping in the S regions, $A_\tau$, $B_\tau$, and $C_\tau$ in Eqs. (\ref{eq26})-(\ref{eq28}) can be reduced to
\begin{align}
& A_\tau = \cos (k_{ex}^\tau w) \cos (k_{hx}^\tau w) - \frac{\sin (k_{ex}^\tau w) \sin (k_{hx}^\tau w)}{\cos \alpha_e^\tau \cos \alpha_h^\tau}, \label{eq13} \\
& B_\tau = \frac{\sin (k_{ex}^\tau w) \cos (k_{hx}^\tau w)}{\cos \alpha_e^\tau} + \frac{\cos (k_{ex}^\tau w) \sin (k_{hx}^\tau w)}{\cos \alpha_h^\tau}, \label{eq14} \\
& C_\tau = - \sin (k_{ex}^\tau w) \sin (k_{hx}^\tau w) \tan \alpha_e^\tau \tan \alpha_h^\tau, \label{eq15}
\end{align}
similar to the results in MLG \cite{Titov}. The $\phi$-dependent energy levels $\epsilon_{n,s}^\tau$ of ABS can be calculated from Eq. (\ref{eq12}), 
and $n$ denotes the number of ABS.
Then the ABS collectively contribute to the cone-related Josephson supercurrent at finite temperature as
\begin{align}
J^\tau = -\frac{2 e d_y}{\pi \hbar} \sum_{n,s} \int k_e^\tau \tanh(\frac{\epsilon_{n,s}^\tau}{2 k_B T}) \frac{d \epsilon_{n,s}^\tau}{d \phi} \cos \alpha_i d \alpha_i, \label{eq16}
\end{align}
with the incident angle $\alpha_i$ of the electron and the transverse width $d_y$ of S/N/S junction.
The total current is defined as $J=J^+ + J^-$ and the unit of Josephson current is $J_0=2 e d_y k_e^\tau / \pi \hbar$.

\begin{figure}
\includegraphics[width=8.0cm,height=6.0cm]{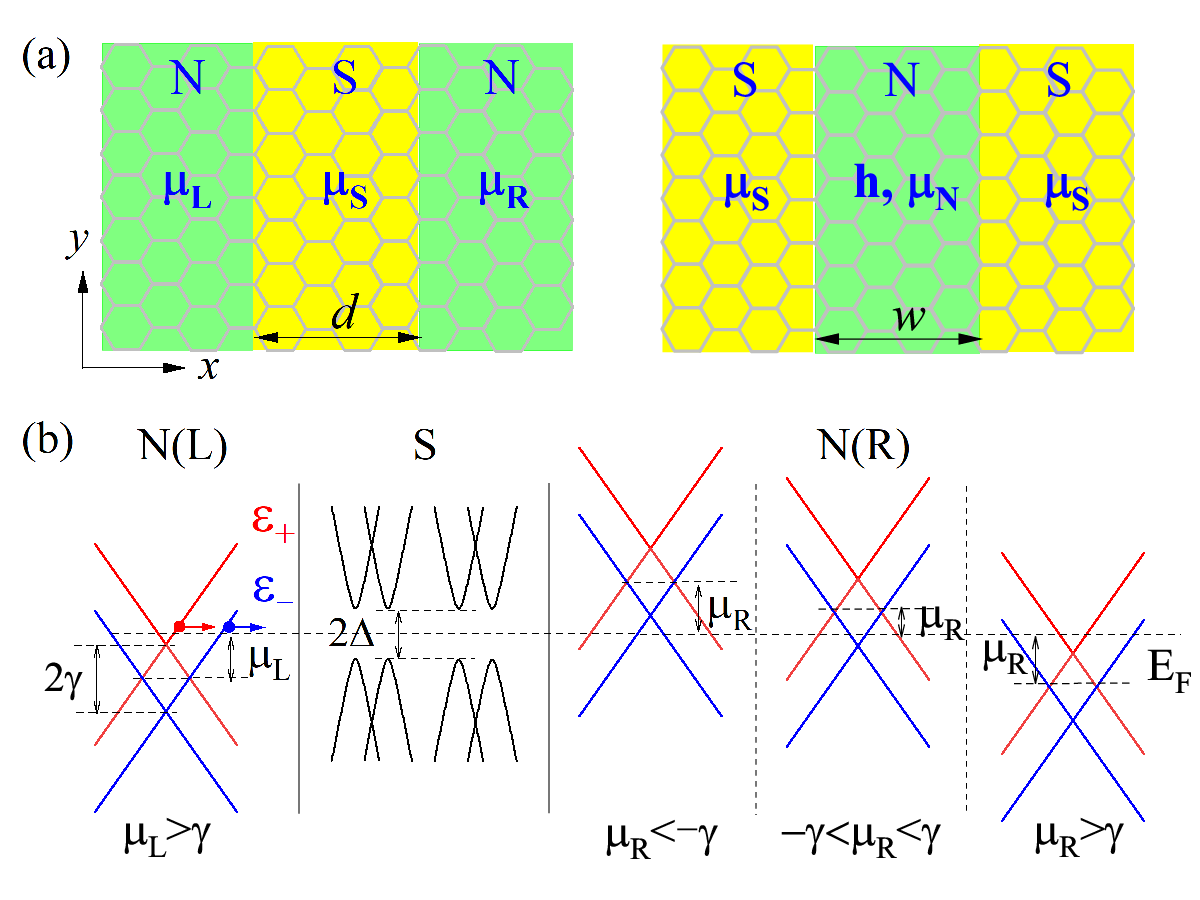}
\caption{(a) Schematic diagrams for the AA-BLG-based N/S/N and S/N/S junctions. $\mu_{L,R,N,S}$ are the chemical potentials and $h$ is the exchange field.
(b) Energy band of the N/S/N junction for different values of $\mu_R$. The red and blue curves denote the upper and lower cones, respectively.}
\end{figure}

\section{Results and discussions}
In this section, we discuss the superconducting lens phenomenon in N/S/N junction, the Josephson current in S/N/S junction, and their dependence on the cone degree of freedom.
The typical superconducting gap $\Delta_0=1.0meV$ and interlayer coupling strength $\gamma  \approx 200meV$ are chosen.
The superconducting coherence length is $\xi = \hbar v_F / \Delta_0 \approx 660nm$ in the superconducting AA-BLG.
For N/S/N junction discussed in Sec. III A, its width $d$ is comparable to $\xi$ and the exchange field is not considered ($h=0$).
For S/N/S junction in Sec. III B, the width $w$ satisfies $w \leq 60nm << \xi$, since short junction is the most relevant experimentally.

\subsection{Superconducting lens}
There exist four scattering processes in N/S/N junction, including normal reflection, local AR, ET, and CAR.
The properties of normal reflection and local AR closely resemble the ones in a single N/S junction \cite{WTLu}.
Here, we mainly focus on ET and CAR in the following discussion, which correspond to the transmission as an electron and a hole, respectively.
In analogy with light refraction, ballistic electron may undergo refraction when it tunnels through a junction.
The negative refraction of MLG proves advantageous for engineering electronic lenses \cite{Cheianov}.
Thus, some interesting lens behaviors for electrons and holes are anticipated when an electron is emitted from different cones in the AA-BLG-based N/S/N junction.

Considering an incident electron beam from a point source in the left N lead, the conservation of the transverse wave vector $k_y$ gives rise to the Snell's law between the incident angle $\alpha_i$ and the refraction angles $\alpha_{e,h}^\tau$ of transmitted electrons and holes,
\begin{align}
k_{eL}^\tau \sin \alpha_i = k_{eR}^\tau \sin \alpha_e^\tau = k_{hR}^\tau \sin \alpha_h^\tau. \label{eq17}
\end{align}
In the limit $\mu_{L,R} \pm \tau \gamma \gg \Delta, \epsilon$, the relative refractive indexes are defined as
\begin{align}
& n_e^\tau = \frac{\sin \alpha_e^\tau}{\sin \alpha_i} = \frac{k_{eL}^\tau}{k_{eR}^\tau} \approx \frac{\mu_L - \tau \gamma}{ \mu_R - \tau \gamma}, \label{eq18} \\
& n_h^\tau = \frac{\sin \alpha_h^\tau}{\sin \alpha_i} = \frac{k_{eL}^\tau}{k_{hR}^\tau} \approx \frac{\mu_L - \tau \gamma}{ - \mu_R + \tau \gamma}, \label{eq19}
\end{align}
with $n_e^\tau=-n_h^\tau$, determining the refraction behaviors of electron and hole for both cones.
Obviously, the sign of refractive indexes $n_{e,h}^\tau$ strongly depends on the interlayer coupling $\gamma$ and chemical potentials $\mu_{L,R}$.
The positive, zero, and negative $n_{e,h}^\tau$ suggest that the transmitted electrons and holes are diverged, collimated, and focused, respectively.
Furthermore, the positive or negative $k_{hR}^\tau$ indicates whether the transmitted holes belong to the conduction or valence band, as does the incident/transmitted electrons with $k_{eL/R}^\tau$.
Without loss of generality, we assume $\mu_L >\gamma$, i.e., both $k_{eL}^+$ and $k_{eL}^-$ keep positive, meaning that the incident electrons are from conduction band for both cones [see Fig. 1(b)].
Note that we assume a heavily doped superconductor, i.e. the chemical potential $\mu_S$ is much larger than $\gamma$, $\Delta$ and the energy $\epsilon$. 
Then the wave vector $k_S$ in S region is much larger than wave vectors $k_{eL}^\tau$, $k_{eR}^\tau$, $k_{hR}^\tau$ in N regions. As a result, the refraction angle $\alpha_S^\tau$ in S region are very small, and the trajectories of electron and hole-like excitations inside S region are approximately in the $x$ direction.
The lensing effect of N/S/N junction with no doping in superconductor is discussed in Appendix C.

\begin{figure}
\includegraphics[width=8.0cm,height=7.0cm]{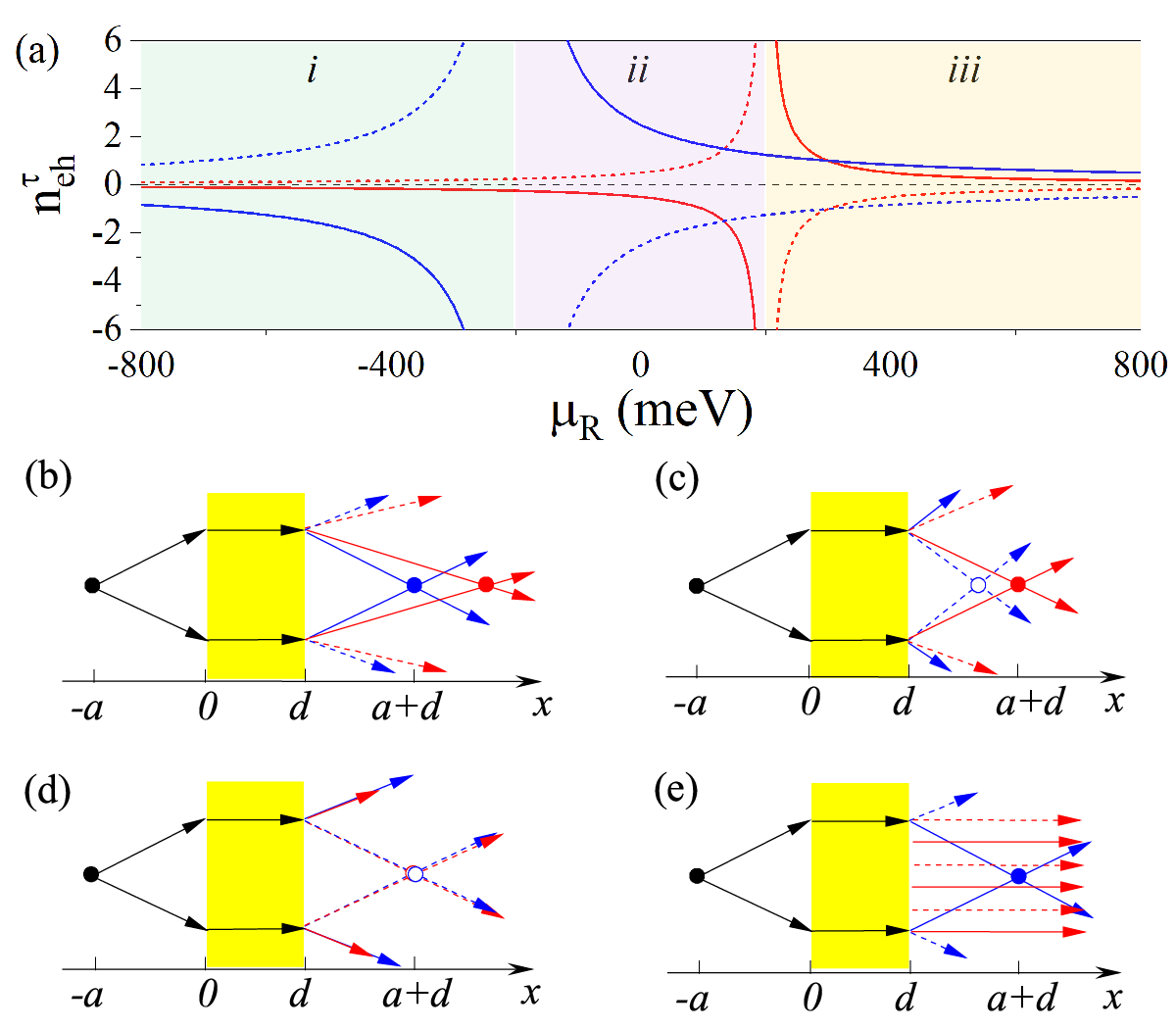}
\caption{ (a) Refractive indexes of the transmitted electrons (solid curve) and holes (dashed curve) from upper cone (red curve) and lower cone (blue curve) through the N/S/N junction when $\mu_L=300 meV$, $\epsilon=0.5 meV$.
[(b)-(e)] Sketch for refractions of the transmitted electrons and holes at (b) $\mu_L=300 meV$, $\mu_R=-700meV$; (c) $\mu_L=300 meV$, $\mu_R=100meV$; (d) $\mu_L=300 meV$, $\mu_R=300meV$; and (e) $\mu_L=201 meV$, $\mu_R=-600meV$.
The red (blue) solid and dashed arrows are for electrons and holes from the upper (lower) cone, respectively.
The black arrows are for the incident electrons from both cones.
Note that the normal reflection and local AR processes are not exhibited.}
\end{figure}

Figure 2(a) shows the cone-dependent refractive indexes $n_{e,h}^\tau$ as a function of potential $\mu_R$ when an incident electron beam is emitted from a point source ($-a$, $0$) in the left N lead.
We can see that four modes with different refractive indexes appear in the right N lead.
Depending on the sign of $n_{e,h}^\tau$, Fig. 2(a) is divided into three regions labeled as $i$, $ii$, and $iii$, respectively, corresponding to the three cases $\mu_R < -\gamma$, $-\gamma < \mu_R < \gamma$, and $\mu_R > \gamma$ of energy band in Fig. 1(b).
The refraction process exhibits the following characteristics:

(1) First, in the $i$ region of Fig. 2(a) at $\mu_R < -\gamma$, for both cones, the holes undergo positive refraction while the electrons undergo negative refraction, implying that the transmitted electrons and holes are focused and diverged, respectively.
Because of interlayer coupling $\gamma$, the refractive index $|n_{e,h}^-|$ of lower cone is greater than $|n_{e,h}^+|$ 
of upper cone when potential $\mu_L > \gamma$.
As a result, the $\tau=+1$ electrons are focused farther away from the junction, but the $\tau=-1$ electrons are focused closer to the junction.
When $\mu_L=300 meV$, $\mu_R=-700meV$ in Fig. 2(b) as an instance, the refractive indexes are $n_e^-=-1$ and $n_e^+=-0.11$, so the $\tau \approx -1$ electrons could be perfectly focused at ($a+d$, $0$), while the $\tau=+1$ electrons will form a pair of caustics that coalesce in a cusp at ($x_e^+ +d$, $0$) and $x_{e,h}^\tau=a/|n_{e,h}^\tau|$.
The caustics $y_{e,h}^\tau (x)$ can be expressed as \cite{Cheianov}:
\begin{align}
& y_{e,h}^\tau (x) = \pm \sqrt{ sgn(1+n_{e,h}^\tau) \frac{(n_{e,h}^\tau)^2 (x^{2/3}-(x_{e,h}^\tau)^{2/3})^3}{(n_{e,h}^\tau)^2-1} }. \label{eqR1}
\end{align}
Note that the CAR process is specular in this case, since the incident electron and transmitted hole occur in the conduction and valence bands, respectively.

(2) As $\mu_R$ increasing to $-\gamma < \mu_R < \gamma$ in the $ii$ region, the refraction of $\tau=-1$ holes becomes negative from positive and the $\tau=+1$ holes still remain positive refraction.
One can clearly see that the two negative refractions $n_e^+$, $n_h^-$ are sensitive to the potential $\mu_R$, and so are the positive refractions $n_e^-$, $n_h^+$.
When $n_e^+=n_h^-$ or $n_e^-=n_h^+$, one may get $\mu_R \approx \gamma^2 / \mu_L$ from Eqs. (\ref{eq18}) and (\ref{eq19}).
Therefore, at $\mu_R = \gamma^2 / \mu_L$, the $\tau=-1$ holes and $\tau=+1$ electrons will be focused at the same region in the right N lead.
Furthermore, one may get $n_e^+=-1$ and $n_h^-=-5/3$ when $\mu_L=300 meV$ and $\mu_R=100meV$, as shown in Fig. 2(c), suggesting that the $\tau=+1$ electrons are focused at the focal point ($a+d$, $0$), while $\tau=-1$ holes form a cusp at ($3a/5 +d$, $0$) [see Fig. 3(c) in the following].

(3) When $\mu_R$ further increases to $\mu_R > \gamma$ in the $iii$ region, the refractive indexes $n_{e,h}^+$ of upper cone change their signs.
Contrary to that in the $i$ region, for both cones, the electrons undergo positive refraction, while the holes are negatively refracted and focused, corresponding to the normal CAR scattering.
The relation $n_h^+=n_h^-$ or $n_e^+=n_e^-$ gives rise to $\mu_R \approx \mu_L$.
When $\mu_R = \mu_L$, we can get $n_h^+=n_h^- \approx -1$ and $n_e^+=n_e^- \approx 1$, 
suggesting that both holes are focused on the symmetric point ($a+d$, $0$) of the source ($-a$, $0$), while both electrons exactly form a virtual focal spot at ($d-a$, $0$) [see Fig. 2(d) at $\mu_R = \mu_L=300meV$].
At $\gamma < \mu_R < \mu_L$ ($\mu_R > \mu_L$), the refractive indexes $|n_{e,h}^+|$ are larger (smaller) than $|n_{e,h}^-|$, 
so from Eq. (\ref{eqR1}) we can conclude that the caustics of $\tau=+1$ holes are closer to (farther away from) the junction than the ones of $\tau=-1$ holes.

(4) In addition, Fig. 2(a) illustrates that for a large negative $\mu_R$, such as $\mu_R=-600meV$, the refractive indexes $n_{h,e}^+ \approx \pm 0.12$ for upper cone at $\mu_L=300meV$ are exceedingly small.
As we approach the limit $\mu_L \rightarrow \gamma$, such as $\mu_L=201meV$ and $\mu_R=-600meV$, $n_{h,e}^+ \approx \pm 0.001$ for upper cone trend to zero, while $n_{h,e}^- \approx \pm 1$ remain significant for lower cone.
As a consequence, we achieve a unique scenario wherein the electron collimation and hole collimation occur for upper cone due to their negligible refractive indexes.
Simultaneously, for the lower cone, the electrons converge at a symmetric point ($a+d$, $0$) of the source, while the virtual focal spot for holes manifests at ($d-a$, $0$).

Therefore, the N/S/N junction can be used as an effective superconducting lens to achieve the focusing and collimation of electrons and holes depending on the cone index in AA-BLG.
Although we set $\mu_L > \gamma$ in Fig. 2, when $\mu_L$ takes other values, the cone-dependent incident electrons from conduction band or valence band would have similar lens effects.

\begin{figure}
\includegraphics[width=8.0cm,height=8.0cm]{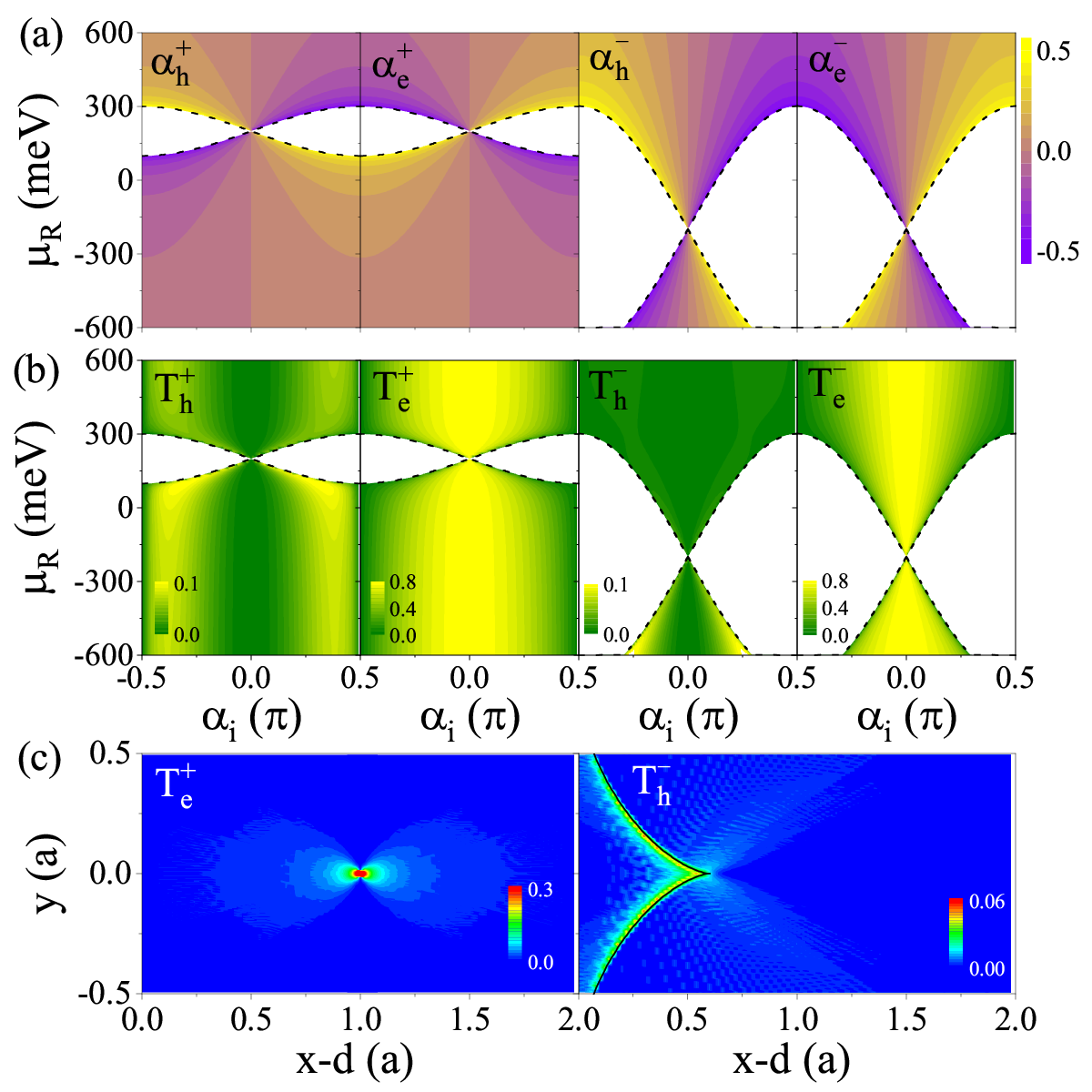}
\caption{ [(a) and (b)] Contour plots for (a) refraction angles $\alpha_{e,h}^\tau$ and (b) transmission probabilities $T_{e,h}^\tau$ in the ($\alpha_i$, $\mu_R$) plane at $\mu_L=300 meV$. The unit of $\alpha_{e,h}^\tau$ is $\pi$ in (a). The white patterns represent the evanescent transmitted modes. 
(c) Contour plots for transmission probabilities $T_e^+$, $T_h^-$ in ($x$, $y$) space when $\mu_L=300 meV$ and $\mu_R=100 meV$.
Other parameters are $\epsilon=0.5 meV$ and $d=0.5 \xi=330 nm$.}
\end{figure}

To further understand the refraction behavior, we calculate the refraction angles and transmission probabilities in the ($\alpha_i$, $\mu_R$) plane, as shown in Figs. 3(a) and 3(b).
The refraction angles take the form as $\alpha_e^\tau = \arcsin (k_{eL}^\tau \sin \alpha_i / k_{eR}^\tau)$ for electrons and $\alpha_h^\tau = \arcsin (k_{eL}^\tau \sin \alpha_i / k_{hR}^\tau)$ for holes.
From Fig. 3(a) one may find that the signs of four types of refraction angles can be adjusted by $\mu_R$.
All refraction angles $\alpha_{e,h}^\tau$ are antisymmetric with respect to $\alpha_i=0$, at which $\alpha_{e,h}^\tau$ are zero.
Setting $\alpha_{e,h}^\tau=\pi/2$ can yield the critical values of incident angles defined by $\alpha_{ec}^\tau = \arcsin (k_{eR}^\tau / k_{eL}^\tau)$, 
$\alpha_{hc}^\tau=\arcsin (k_{hR}^\tau / k_{eL}^\tau)$, 
as marked by the dashed curves in Figs. 3(a) and 3(b), and the white areas suggest a total reflection.
In the vicinity of $\mu_R=\tau \gamma$, the allowable incident angles for electrons and holes gradually decrease.
The transmission probabilities $T_{e,h}^\tau$ are symmetric around $\alpha_i=0$ [see Fig. 3(b)].
For both cones, the electron transmissions $T_e^\pm$ reach their maximum at $\alpha_i=0$, while the maximums of hole transmissions $T_h^\pm$ strongly depend on $\alpha_i$ and $\mu_R$.

When $\mu_L=300meV$ and $\mu_R=100meV$, taking $n_e^+=-1$ and $n_h^-=-5/3$ as an instance, Fig. 3(c) displays the transmission probabilities $T_e^+$ and $T_h^-$ in ($x, y$) space calculated by Green's function method \cite{Moghaddam, Sun5}, corresponding to the semiclassical trajectory plotted in Fig. 2(c).
The black curves in the right column of Fig. 3(c) are calculated by Eq. (\ref{eqR1}).
One may find that $T_e^+$ has a peak at ($a+d, 0$). 
The peaks of $T_h^-$ form two bright zones suggesting a pair of caustics and they coalesce in a cusp at ($3a/5+d, 0$).
The results of Fig. 3(c) are consistent with those of Fig. 2(c) and Eq. (\ref{eqR1}).

\begin{figure}
\includegraphics[width=8.0cm,height=6.0cm]{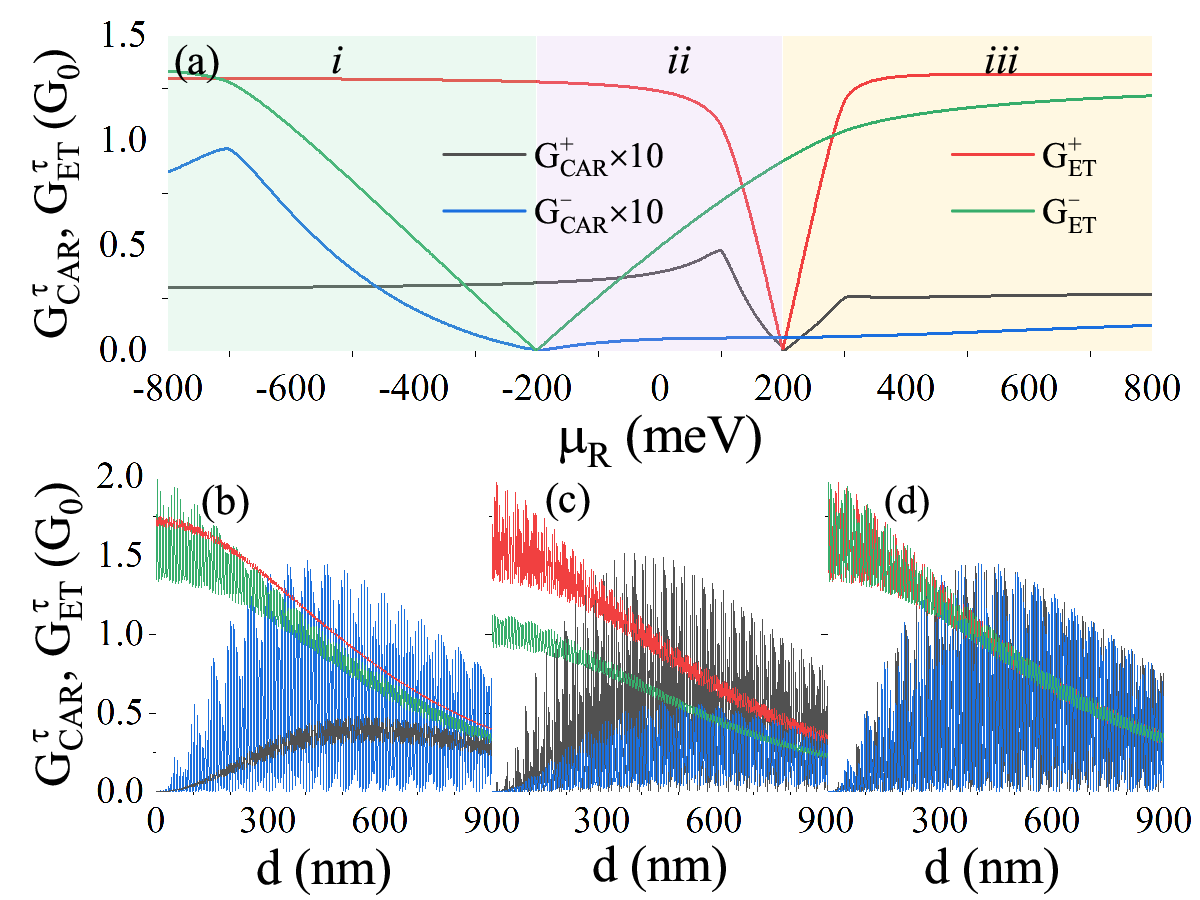}
\caption{ (a) Conductances $G_{CAR}^\tau \times 10$, $G_{ET}^\tau$ versus $\mu_R$ at $d=330 nm$.
[(b)-(d)] Conductances $G_{CAR}^\tau \times 10$, $G_{ET}^\tau$ versus width $d$ when (b) $\mu_R=-700 meV$, (c) $\mu_R=100 meV$, and (d) $\mu_R=300meV$.
Here, $\mu_L=300 meV$ and $\epsilon=0.5meV$.}
\end{figure}

Above, we show that the superconducting lens effect is caused by CAR and ET scatterings.
As a measurable quantity, it is necessary to discuss the property of conductances $G_{CAR}^\tau$ and $G_{ET}^\tau$ contributed by CAR and ET.
According to the focusing characteristics shown in Figs. 2(b)-2(e), the drains positioned strategically at the focal spots allow measurement of the cone-dependent nonlocal differential conductance $G^\tau = G_{CAR}^\tau- G_{ET}^\tau$ carried by electrons and holes.
Fig. 4(a) shows the dependence of conductances $G_{CAR}^\tau$, $G_{ET}^\tau$ on the potential $\mu_R$.
With the increase of $\mu_R$, $G_{CAR}^-$ and $G_{ET}^\pm$ decrease to zero before gradually increasing, while the change of $G_{CAR}^+$ is complex where two peaks appear near the nadir [see Fig. 4(a)], which can be understood by the transmission in Fig. 3(b).
Zero conductances occur at $\mu_R=\tau \gamma \mp \epsilon$, corresponding to Dirac points in the band of electron and hole, where the group velocity is zero and the sign of refractive indexes is changed [see Figs. 1(b) and 2(a)].
Dramatically, the conductances in Fig. 4(a) can be divided into three regions, in accordance with the three regions in Fig. 2(a).
Based on Figs. 2(b)-2(d), we conclude that in the $i$ region of Fig. 4(a), the two cone-dependent conductances $G_{ET}^\pm$ contributed by electrons can be measured by setting the drains at the corresponding focal spots or near the cusps.
Likewise, at the focal spots or near the cusps, we can measure $G_{ET}^+$ and $G_{CAR}^-$ in the $ii$ region, as well as $G_{CAR}^\pm$ in the $iii$ region.

Specifically, Figs. 4(b)-4(d) present the oscillation behavior of conductances $G_{CAR}^\tau$, $G_{ET}^\tau$ as a function of the width $d$ at potential $\mu_R=-700meV$, $100meV$, and $300meV$, corresponding to the refractive indexes $n_e^- = -1$, $n_e^+ = -1$, and $n_h^\pm = -1$ in Figs. 2(b)-2(d), respectively.
As the width $d$ increases, $G_{ET}^\pm$ exhibit damped oscillation, while $G_{CAR}^\pm$ show a typical oscillation which initially increase, reaching a maximum value near the superconducting coherence length $\xi$, and then gradually decrease, consistent with previous reports \cite{Cayssol, Sun, addr1, Linder, Pandey, Sun2}.
This means that at the focal spot ($a+d$, $0$), one may detect the large conductances $G_{ET}^\pm$ under a narrow width, and the large $G_{CAR}^\pm$ when $d$ trends to $\xi$.
Note that the conductances $G_{CAR}^\pm$ are relatively small at $\mu_L=300 meV$, due to the bad mismatch of wave functions between the incident electron and transmitted hole.
In the vicinity of the Dirac point of energy band with $\mu_{L,R} \rightarrow \tau \gamma$, $G_{CAR}^\tau$ and $G_{ET}^\tau$ can be effectively controlled. 
For appropriate values of $\mu_{L,R}$, such as $\mu_L=200.5 meV$ and $\mu_R=199.5 meV$, $G_{CAR}^+$ may be significantly enhanced, but $G_{ET}^+$ will be completely suppressed because its group velocity is zero at $\mu_R= \gamma - \epsilon=199.5 meV$ \cite{Cayssol}.

\subsection{Josephson effect}

In the S/N/S junction, the Josephson supercurrent could occur due to the appearance of ABS.
Josephson effect has been extensively studied in various systems including MLG \cite{Titov, Sun3, Linder2, Sun4, Wakamura, Pellegrino, Wang}, Bernal BLG \cite{Xie, Park, Rout}, and twisted BLG \cite{Xie2, Alvarado, Diez, Sainz, Hu}.
Herein we concentrate on the cone-dependent Josephson effect in AA-BLG.
From Eq. (\ref{eq11}) 
one can infer that supercurrent is permitted only within intracone and it is prohibited in intercone process.

\begin{figure}
\includegraphics[width=8.0cm,height=6.0cm]{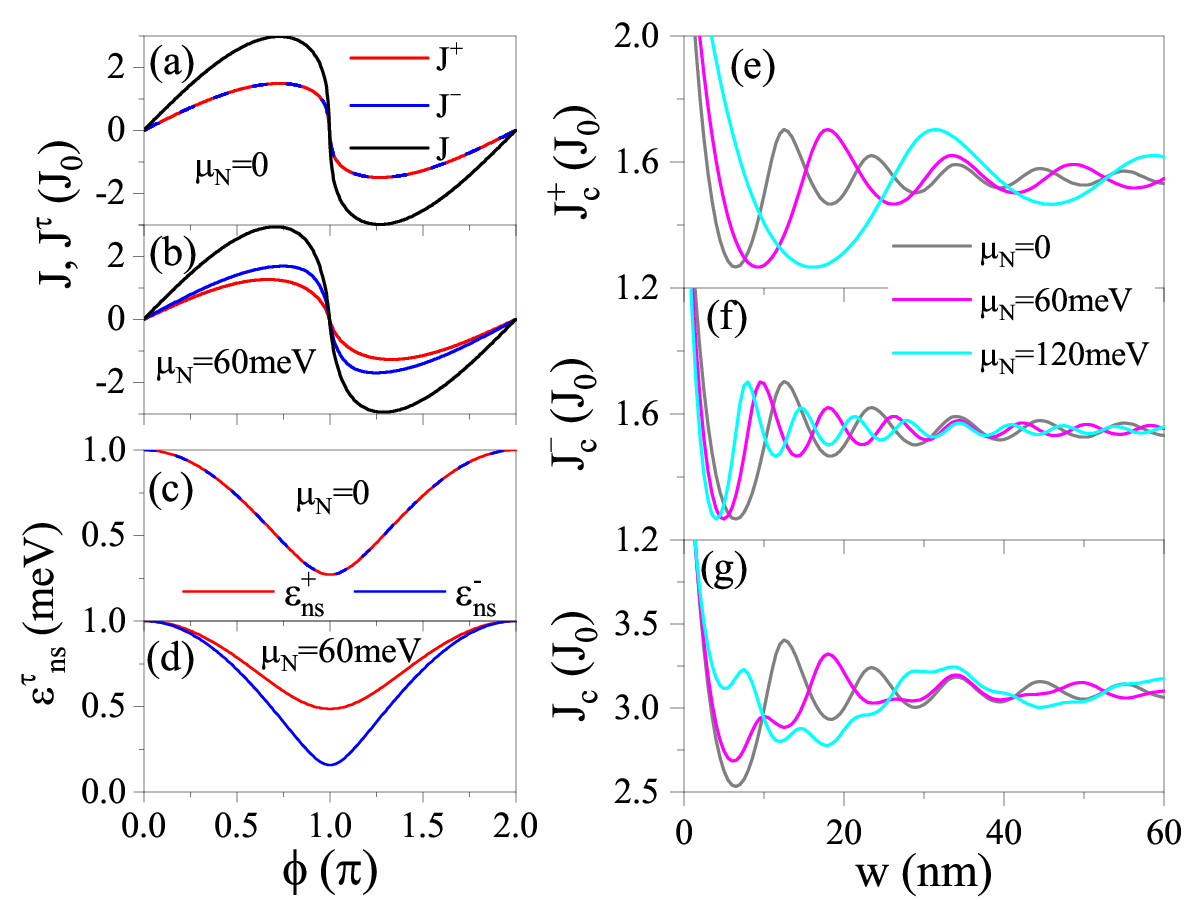}
\caption{ [(a) and (b)] Current-phase relationship in S/N/S junction at (a) $\mu_N=0$ and (b) $\mu_N=60meV$ with $w=10nm$.
(c) and (d) are the ABS levels $\epsilon_{n,s}^\tau$ for both cones at $\alpha_e^\tau=\pi/6$, corresponding to the current-phase relationships in (a) and (b), respectively.
[(e)-(g)] Critical supercurrent versus the width $w$. Other parameters are $h=0$ and $T=0$.}
\end{figure}

First, we discuss the supercurrent in Fig. 5 without exchange field ($h=0$).
In the absence of $\mu_N$ in Fig. 5(a), the currents $J^\pm$ of the two cones are the same due to their degenerate ABS levels [see Fig. 5(c)] and the current-ABS levels relation in Eq. (\ref{eq16}).
$J(\phi)$ is the sum of $J(\phi)^+$ and $J(\phi)^-$.
When $\mu_N \neq 0$ in Fig. 5(b), the currents $J^\pm$ are no longer equal since the dispersion of their ABS levels becomes different.
The larger slope $d \epsilon_{n,s}^- / d \phi$ of $\phi$-dependent ABS level for lower cone [see Fig. 5(d)] causes a larger current than the upper cone.
The current-phase relation is nonsinusoidal, where the positions of the maxima of the currents are shifted to $\phi=\pi$ since the slope $d \epsilon_{n,s}^\tau / d \phi$ is large near $\phi=\pi$.
Because of the time-reversal symmetry, the currents $J^\pm$ of both cones and the total current $J$ are approximately proportional to the phase difference $\phi$, which correspond to the $0$ phase of the S/N/S junction. 
The Josephson junction still keeps $0$ phase for large potential $\mu_N$.
Figs. 5(e)-5(g) plot the critical currents (e) $J_c^+$, (f) $J_c^-$, and (g) $J_c$ as a function of junction width $w$ for different values of the potential $\mu_N$.
$J_c^\tau$ and $J_c$ are the maximum/minimum values of $J^\tau (\phi)$ and $J(\phi)$, respectively, in the interval $0<\phi<\pi$. 
It can be seen that all the critical currents oscillate damply with the width $w$ yet possess distinct oscillation periods.
The oscillation originates from transmission resonance of the quasiparticles, and the cone-dependent oscillation period $T_w^\tau$ can be given approximately by
\begin{align}
T_w^\tau \approx \frac{\pi}{|k_e^\tau|} \approx \frac{\pi \hbar v_F}{|\mu_N - \tau \gamma|}. \label{eq20}
\end{align}
Equation (\ref{eq20}) indicates that due to interlayer coupling $\gamma$, with the increase of $\mu_N$, the oscillation period $T_w^-$ for critical current $J_c^-$ decreases, while $T_w^+$ for $J_c^+$ increases before $\mu_N=\gamma$ and then decreases, as displayed in Figs. 5(e) and 5(f).
Owing to the diverse oscillation periods $T_w^\pm$, the total critical current $J_c$ oscillates irregularly with $w$ [see Fig. 5(g)].

\begin{figure}
\includegraphics[width=8.0cm,height=6.0cm]{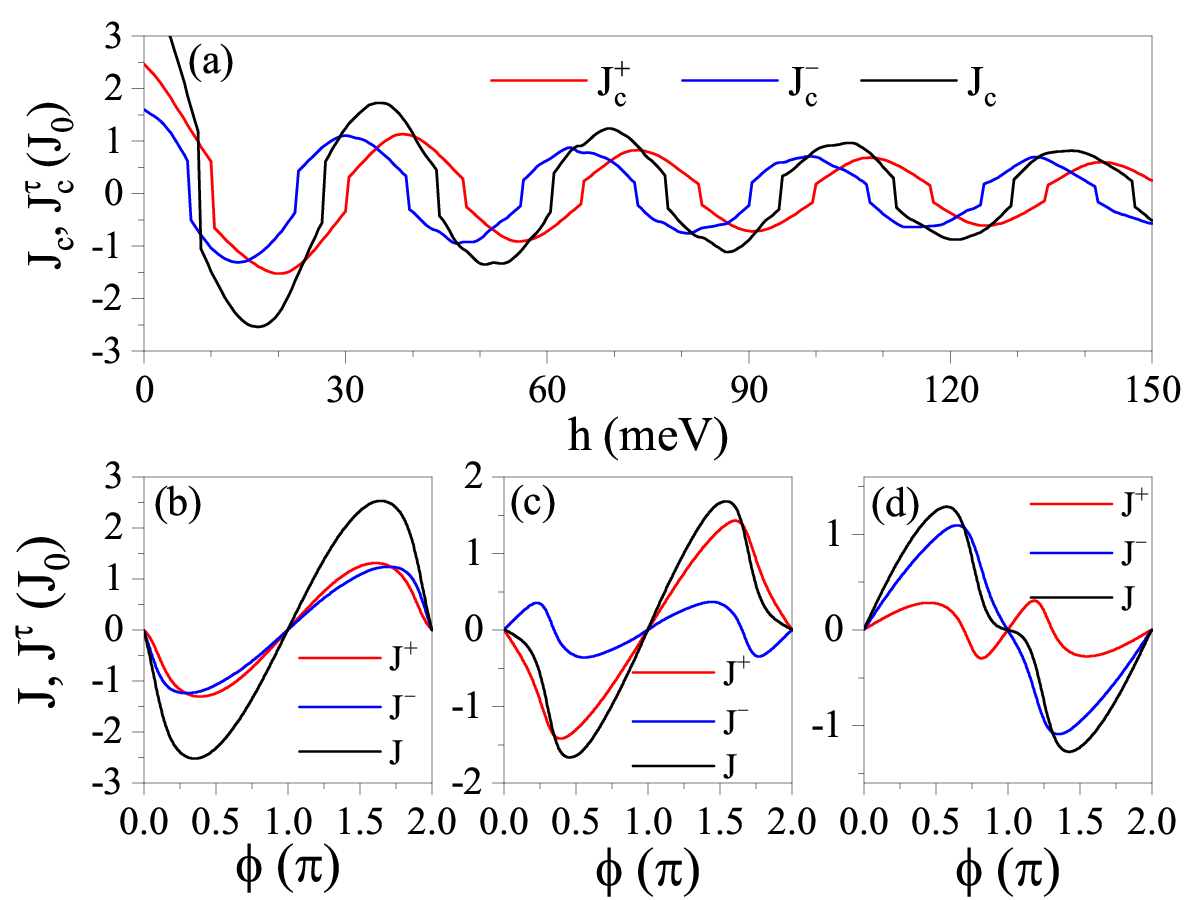}
\caption{ (a) Critical supercurrent versus the exchange field $h$.
[(b)-(d)] Current-phase relationship at (b) $h=16meV$, (c) $h=22.5meV$, and (d) $h=30.5meV$. Other parameters are $\mu_N=200meV$, $w=60nm$, and $T=0.1T_c$.}
\end{figure}

The presence of ferromagnetic exchange field $h$ in the center N region 
would break the time-reversal symmetry of the system, and so the supercurrent reversal is expected.
Figure 6(a) exhibits the change of critical currents $J_c^\pm$, $J_c$ with $h$ at a low temperature $T=0.1T_c$.
It is evident that as $h$ increases, all critical currents oscillate between positive value and negative value, meaning that the Josephson junction undergoes a transition from $0$ phase to $\pi$ phase.
Importantly, the $0-\pi$ transition strongly depends on the cone index.
Because of $J_c^\tau \propto \cos(k_{ex}^\tau w)$, the interlayer coupling $\gamma$ results in a stable phase difference of one quarter period between $J_c^+$ and $J_c^-$. To be precise, $J_c^-$ leads $J_c^+$ by one quarter period, and $J_c$ leads $J_c^+$ by one eighth period.
All three share an identical oscillation period, which can be approximated as
\begin{align}
T_h \approx \frac{\pi \hbar v_F}{w}. \label{eq21}
\end{align}
The period $T_h$ can be modulated by the junction width $w$.
In consequence, when the supercurrent of a certain cone is in a stable $0$ phase or $\pi$ phase, 
its opposite cone may experience a $0-\pi$ transition.
The total critical current $J_c$ is determined by the competition between the critical currents $J_c^\pm$.
Taking $h=16meV$, $22.5meV$, and $30.5meV$ for instance, the current-phase relationship is depicted in Figs. 6(b)-6(d).
At $h=16meV$ in Fig. 6(b), both currents $J^\pm$ are in $\pi$ phases, so $J$ is the same.
When $h$ increases to $22.5meV$ in Fig. 6(c), $J^-$ becomes a coexistence of $0$ and $\pi$ phases, suggesting a residual value of $J^-$ at $0-\pi$ transition, similar to that observed in MLG \cite{Linder2}. At the same time, $J^+$ remains in $\pi$ phase, and $J$ is also in $\pi$ phase.
Oppositely, as $h$ increases to $30.5meV$, the coexistence of $0$ and $\pi$ phases emerges in $J^+$, while $J^-$ and $J$ are in the stable $0$ phase [see Fig. 6(d)].
The $0-\pi$ transition still holds at higher values of $h$.
In addition, the appearance of $h$ lifts the spin degeneracy of ABS levels and the number of various ABS levels increases. In consequence, the nonsinusoidal form of current-phase relation becomes more prominent, resulting in a residual supercurrent at the transition point.
These results prove that the cone-dependent $0-\pi$ transition can be achieved and controlled in AA-BLG.

\section{Conclusion}
In summary, we studied the Andreev scattering in AA-BLG, which exhibits interesting properties distinct from MLG and Bernal BLG, due to the unique band structure and interlayer coupling of AA-BLG.
By adjusting the potential relative to interlayer coupling, the N/S/N junction can function as a cone-related superconducting lens to focus holes and electrons in different spatial regions.
In the S/N/S junction, the critical currents of the two cones exhibit different oscillation behaviors.
There exists a phase difference of one quarter period between these critical currents, leading to a cone-dependent $0-\pi$ transition.
The findings are helpful for the design of superconducting electronic devices based on AA-BLG.

\section*{Acknowledgments}
This work was supported by the National Natural Science Foundation of China (Grants No. 12374034, No. 12474045, No. 11974153, and No. 11921005), the Innovation Program for Quantum Science and Technology (Grant No. 2021ZD0302403), and the National Key R and D Program of China (Grant Nos. 2024YFA1409002).

\appendix


\section{Eigenstates of BdG equation}

By solving the BdG equation (\ref{eq2}), 
for given energy $\epsilon$ and transverse wave vector $k_y$, the eigenstates in the N region of considered junctions are
\begin{align}
\phi_{\tau e}^{\pm} (x) = \frac{1}{\sqrt{\cos \alpha_e^\tau}} \left(\begin{array}{cc} \tau e^{\mp i \alpha_e^\tau/2}  \\  \pm \tau e^{\pm i \alpha_e^\tau/2}  \\  e^{\mp i \alpha_e^\tau/2}   \\  \pm e^{\pm i \alpha_e^\tau/2}  \\  O  \end{array}\right) e^{\pm i k_{ex}^\tau x}, \label{eq22} \\
\phi_{\tau h}^{\pm} (x) = \frac{1}{\sqrt{\cos \alpha_h^\tau}} \left(\begin{array}{cc} O  \\  \tau e^{\mp i \alpha_h^\tau/2}  \\  \mp \tau e^{\pm i \alpha_h^\tau/2}  \\  e^{\mp i \alpha_h^\tau/2}   \\  \mp e^{\pm i \alpha_h^\tau/2}  \end{array}\right) e^{\pm i k_{hx}^\tau x}, \label{eq23}
\end{align}
with the null matrix $O = (0, 0, 0, 0)^T$.
The state $\phi_{\tau e}^{\pm}$ (or $\phi_{\tau h}^{\pm}$) represents the electron (or hole) propagating in the $\pm x$ direction for the cone $\tau$.
The wave vectors for electron and hole are $k_e^\tau=(\epsilon + \mu_N + s h - \tau \gamma) / \hbar v_F$ and $k_h^\tau=(\epsilon - \mu_N + s h + \tau \gamma) / \hbar v_F$, respectively.
$\alpha_e^\tau=\arcsin(k_y/k_e^\tau)$ is the incident angle and also the refraction angle of electron.
$\alpha_h^\tau=\arcsin(k_y/k_h^\tau)$ is the refraction angle of the transmitted hole.
$k_{ex}^\tau=k_e^\tau \cos \alpha_e^\tau$ and $k_{hx}^\tau=k_h^\tau \cos \alpha_h^\tau$ are the corresponding longitudinal wave vectors.
The factors $1/\sqrt{\cos \alpha_{e,h}^\tau}$ in Eqs. (\ref{eq22}) and (\ref{eq23}) ensure the probability current conservation.

In the limit $\mu_S + \tau \gamma \gg \Delta, \epsilon$, the eigenstates in the S region can be written as
\begin{align}
\psi_{\tau S1}^{\pm} (x) = \left(\begin{array}{cc}  e^{\mp i \beta}  \\  \pm e^{\pm i \zeta_\tau \mp i \beta}  \\  - \tau e^{\mp i \beta}   \\  \mp \tau e^{\pm i \zeta_\tau \mp i \beta}  \\  e^{- i \phi}    \\  \pm e^{\pm i \zeta_\tau - i \phi}  \\  - \tau e^{- i \phi}  \\  \mp \tau e^{\pm i \zeta_\tau - i \phi}  \end{array}\right) e^{\pm i q_\tau x + \kappa_\tau x}, \label{eq24} \\
\psi_{\tau S2}^{\pm} (x) = \left(\begin{array}{cc}  e^{\pm i \beta}  \\  \pm e^{\mp i \zeta_\tau \pm i \beta}  \\  - \tau e^{\pm i \beta}   \\  \mp \tau e^{\mp i \zeta_\tau \pm i \beta}  \\  e^{- i \phi}    \\  \pm e^{\mp i \zeta_\tau - i \phi}  \\  - \tau e^{- i \phi}  \\  \mp \tau e^{\mp i \zeta_\tau - i \phi}  \end{array}\right) e^{\pm i q_\tau x - \kappa_\tau x}, \label{eq25}
\end{align}
where the scattering angle $\zeta_\tau=\arcsin [\hbar v_F k_y / (\mu_S - \tau \gamma)]$, $q_\tau=\sqrt{(\mu_S-\tau \gamma)^2/(\hbar v_F)^2 - k_y^2}$, and $\kappa_\tau=(\mu_S-\tau \gamma) \Delta \sin \beta / [(\hbar v_F)^2 q_\tau]$.
The parameter $\beta=\arcsin (\epsilon / \Delta)$ at $\epsilon < \Delta$, otherwise, $\beta=-i arcosh (\epsilon / \Delta)$.
For the cone $\tau$ at $\epsilon > \Delta$, the state $\psi_{\tau S1}^{\pm}$ represents the quasihole/quasielectron propagating in the $-x$ direction, while $\psi_{\tau S2}^{\pm}$ represents the quasielectron/quasihole propagating in the $+x$ direction.
These states are coherent superpositions of electron and hole excitations in the S region when $\epsilon < \Delta$.

\section{Andreev bound states}
Based on Eqs. (\ref{eq10}) and (\ref{eq22})-(\ref{eq25}), 
the parameters $A_\tau$, $B_\tau$, and $C_\tau$ in Eq. (\ref{eq11}) can be obtained, which take the form
\begin{widetext}
\begin{align}
A_\tau=& \cos (k_{ex}^\tau w) \cos (k_{hx}^\tau w) - \sin (k_{ex}^\tau w) \sin (k_{hx}^\tau w)  [\frac{1+\sin \zeta_\tau (\sin \alpha_h^\tau - \sin \alpha_e^\tau)}{\cos \alpha_e^\tau \cos \alpha_h^\tau \cos^2 \zeta_\tau} - \tan \alpha_e^\tau \tan \alpha_h^\tau \tan^2 \zeta_\tau], \label{eq26} \\
B_\tau =& \sin (k_{ex}^\tau w) \cos (k_{hx}^\tau w) \frac{1-\sin \alpha_e^\tau \sin \zeta_\tau}{\cos \alpha_e^\tau \cos \zeta_\tau} +  \cos (k_{ex}^\tau w) \sin (k_{hx}^\tau w) \frac{1+\sin \alpha_h^\tau \sin \zeta_\tau}{\cos \alpha_h^\tau \cos \zeta_\tau}, \label{eq27}\\
C_\tau =& \sin (k_{ex}^\tau w) \sin (k_{hx}^\tau w)  \frac{\sin^2 \zeta_\tau - \sin \alpha_e^\tau \sin \alpha_h^\tau + \sin \zeta_\tau (\sin \alpha_h^\tau - \sin \alpha_e^\tau)}{\cos \alpha_e^\tau \cos \alpha_h^\tau \cos^2 \zeta_\tau}. \label{eq28}
\end{align}
\end{widetext}
Recent works proved that the interlayer coupling $\gamma$ 
could be controlled by applying pressure from a scanning tunneling microscopy tip \cite{Yankowitz} and by an interlayer potential difference \cite{Zheng}.
In order to calculate conveniently, it is reasonable to consider a heavily doped superconductor ($\mu_S \gg \mu_N >\hbar v_F k_y $ and 
$\mu_S \gg \gamma$), leading to the limit $\zeta_\tau \rightarrow 0$.
Equations (\ref{eq26})-(\ref{eq28}) thus simplify to Eqs. (\ref{eq13})-(\ref{eq15}).

\section{Superconducting lens at $\mu_S=0$}

\begin{figure}
\includegraphics[width=8.0cm,height=8.0cm]{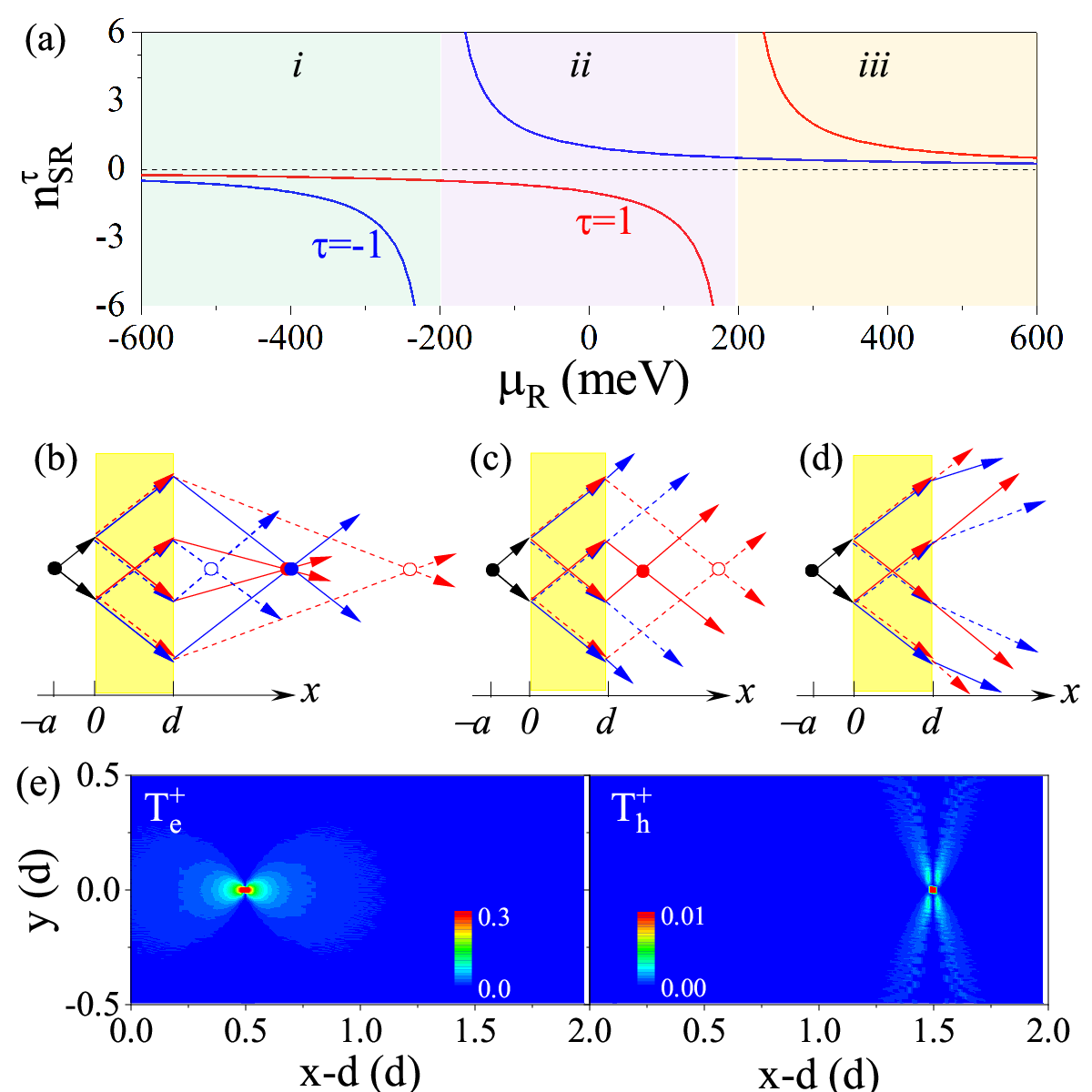}
\caption{(a) Refractive indexes $n_{SR}^\tau$ versus the potential $\mu_R$ in N/S/N junction. 
[(b)-(d)] Sketch for refractions of the transmitted electrons and holes at (b) $\mu_R=-400meV$,  (c) $\mu_R=0$, and  (d) $\mu_R=400meV$. 
The red (blue) solid and dashed arrows are for electrons and holes from the upper (lower) cone, respectively.
The black arrows are for the incident electrons from both cones.
(e) Transmission probabilities $T_e^+$ and $T_h^+$ in ($x$, $y$) space at $\mu_R=0$.
Here, $\mu_L=\mu_S=0$, $\epsilon=0.5 meV$, and $d=330 nm$.}
\end{figure}

When $\mu_S=0$, the refraction in S region would affect the focusing in the right N lead of N/S/N junction.
Specifically, due to $\gamma \gg \Delta, \epsilon$, at $\mu_L=\mu_S=0$, the wave vectors $k_{eL}^\tau \approx -\tau \gamma$, while $k_S \approx \iota \gamma$ with $\iota=\pm 1$ for electron and hole-like excitations.  Then, at the N(L)/S interface, the conservation of $k_y$ would lead to the relative refractive indexes  
\begin{align}
n_{LS}^{\tau,\iota} = \frac{\sin \alpha_S^\tau}{\sin \alpha_i} = \frac{k_{eL}^\tau}{k_S} = \frac{ - \tau}{\iota}. 
\end{align}
Particularly, $n_{LS}^{+,e}=-1$ ($n_{LS}^{+,h}=1$) for electron (hole) from upper cone, while $n_{LS}^{-,e}=1$ ($n_{LS}^{-,h}=-1$) for electron (hole) from lower cone. The semiclassical trajectory of the transmitted electrons and holes at N(L)/S interface is displayed in Figs. 7(b)-7(d). In the right N lead, $k_{eR}^\tau \approx \mu_R - \tau \gamma$ and $k_{hR}^\tau \approx - \mu_R + \tau \gamma \approx -k_{eR}^\tau$. 
Hence, at the S/N(R) interface, the transmitted electrons and holes have the same refractive indexes, 
\begin{align}
n_{SR}^{\tau} = \frac{\sin \alpha_R^\tau}{\sin \alpha_S^\tau} = \frac{\gamma}{\mu_R - \tau \gamma}. 
\end{align}
As shown in Fig. 7(a), the refractive indexes $n_{SR}^{\tau}$ can be controlled by the potential $\mu_R$.
When $\mu_R<-\gamma$, the refractive indexes $n_{SR}^{\tau}$ are negative for both upper and lower cones.
The refractive index $n_{SR}^-$ becomes positive for lower cones at $-\gamma<\mu_R<\gamma$.
Both refractive indexes $n_{SR}^{\tau}$ are positive at $\mu_R>\gamma$.

Here we mainly discuss the ideal case where $\mu_L = \mu_S= 0$ and incident electron is at source ($-a, 0$) with $a=d/2$, as shown in Figs. 7(b)-7(e).
In Fig. 7(b) at $\mu_R=-400meV$, $n_{SR}^- = -1$ but $n_{SR}^+ = -1/3$. In consequence, the hole and electron from lower cone form perfect focal spots at ($3d/2, 0$) and ($5d/2, 0$), respectively, while the electron and hole from upper cone form causticses near the cusps ($5d/2, 0$) and ($9d/2, 0$).
In Fig. 7(c) at $\mu_R=0$, $n_{SR}^+ = -1$ but $n_{SR}^- = 1$, suggesting that the electron and hole from upper cone are focused at ($3d/2, 0$) and ($5d/2, 0$), respectively, and the distance between the two focal spots is the width $d$.
This feature is similar to that in monolayer graphene \cite{Gomez}. 
At the same time, the electron and hole from lower cone can form the virtual focal spots at ($-d/2, 0$) and ($d/2, 0$).
On the contrary, in Fig. 7(d) at $\mu_R=400meV$, $n_{SR}^+ = 1$ and so the hole and electron from upper cone can form the virtual focal spots at ($-d/2, 0$) and ($d/2, 0$).
Fig. 7(e) displays the transmission probabilities $T_e^+$ and $T_h^+$ in ($x, y$) space at $\mu_R=0$, corresponding to the semiclassical trajectory in Fig. 7(c).
We can see that $T_e^+$ and $T_h^+$ have a peak at ($3d/2, 0$) and ($5d/2, 0$), respectively, consistent with the focal spots described in Fig. 7(c).

\end{CJK*}
\end{document}